\documentclass[sigconf]{acmart}
\usepackage{hyperref}
\AtBeginDocument{%
  \providecommand\BibTeX{{%
    \normalfont B\kern-0.5em{\scshape i\kern-0.25em b}\kern-0.8em\TeX}}}

\setcopyright{acmcopyright}
\copyrightyear{2024}
\acmYear{2024}
\acmDOI{TBD}

\acmConference[The Web Conference 2024]{The Web Conference 2024}{May 13--17, 2024}{Singapore}
\acmPrice{15.00}
\acmISBN{978-1-4503-XXXX-X/18/06}

\begin{document}

\title{On-Device Recommender Systems: A Tutorial on \\
The New-Generation Recommendation Paradigm}

\author{Hongzhi Yin}
\affiliation{%
  \institution{The University of Queensland}
  \city{Brisbane}
  \state{QLD}
  \country{Australia}
}
\email{h.yin1@uq.edu.au}

\author{Tong Chen}
\affiliation{%
  \institution{The University of Queensland}
  \city{Brisbane}
  \state{QLD}
  \country{Australia}
  \postcode{4067}
}
\email{tong.chen@uq.edu.au}

\author{Liang Qu}
\affiliation{%
  \institution{The University of Queensland}
  \city{Brisbane}
  \state{QLD}
  \country{Australia}}
\email{liang.qu@uq.edu.au}

\author{Bin Cui}
\affiliation{%
  \institution{Peking University}
  \city{Beijing}
  \country{China}
}
\email{bin.cui@pku.edu.cn}

\renewcommand{\shortauthors}{Hongzhi Yin, et al.}

\begin{abstract}
Given the sheer volume of contemporary e-commerce applications, recommender systems (RSs) have gained significant attention in both academia and industry. However, traditional cloud-based RSs face inevitable challenges, such as resource-intensive computation, reliance on network access, and privacy breaches. In response, a new paradigm called on-device recommender systems (ODRSs) has emerged recently in various industries like Taobao, Google, and Kuaishou. ODRSs unleash the computational capacity of user devices with lightweight recommendation models tailored for resource-constrained environments, enabling real-time inference with users' local data. This tutorial aims to systematically introduce methodologies of ODRSs, including (1) an overview of existing research on ODRSs; (2) a comprehensive taxonomy of ODRSs, where the core technical content to be covered span across three major ODRS research directions, including on-device deployment and inference, on-device training, and privacy/security of ODRSs; (3) limitations and future directions of ODRSs.
This tutorial expects to lay the foundation and spark new insights for follow-up research and applications concerning this new recommendation paradigm.
\end{abstract}





\maketitle

\section{Topic and relevance}
As an indispensable means for web users to counteract information overload, recommender systems (RSs) that can automatically match user interests with relevant items (e.g., products, services, information) have seen a substantial amount of research interest over the last decade. In the digital industry, prosperous enterprise-level applications are projected to drive the global RS market to an unprecedented value of USD \$54 billion by 2030\footnote{\href{https://straitsresearch.com/report/recommendation-engines-market\#:~:text=Market\%20Overview,period\%20(2022-2030).}{https://straitsresearch.com/report/recommendation/}}.

Traditional RSs are subsumed under a fully cloud-based paradigm, where the cloud server trains the RS model with all the user data it hosts and pushes recommendation results to users' personal devices upon request. Though this paradigm enjoys benefits from the ``infinite'' computing power to support sophisticated RS models, some increasingly harsh obstacles are arising in the meantime, including the high resource and energy consumption \cite{wang2020next}, reliance on network access for timeliness \cite{long2023decentralized}, and threat to user privacy \cite{muhammad2020fedfast}, which challenge the sustainability and trustworthiness of cloud-based RSs.

To this end, recent years have witnessed the development of a new yet fast-evolving recommendation paradigm -- on-device recommender systems (ODRSs). Compared with cloud-based RSs, the most significant difference of ODRSs is that users' devices become a key part of the computation on top of their original role of displaying generated recommendations. Typical ODRSs are optimized towards three objectives: (1) on-device deployment and inference that aims to derive a resource-efficient model with minimum accuracy degradation; (2) on-device training and updating that enables the lightweight model to stay up-to-date; and (3) privacy and security mechanisms that respectively keep users and on-device models from malicious attacks. As a result, the heavy cloud-based RS models could be replaced by their lightweight on-device counterparts, such that the inference can be efficiently performed on resource-constrained user devices with locally stored user data \cite{chen2021learning}. In the industry, ODRSs have seen an emerging list of applications, including Taobao's mobile service \cite{gong2020edgerec}, Google's TensorFlow Lite Recommendation API\footnote{\href{https://www.tensorflow.org/lite/examples/recommendation/overview}{https://www.tensorflow.org/lite/examples/recommendation/overview}}, real-time short video recommendation on Kuaishou \cite{gong2022real}, and the built-in recommendation engine in the Brave Web Browser\footnote{\href{https://brave.com/federated-learning/}{https://brave.com/federated-learning/}}.

\textbf{Rationale of The Tutorial.} Given the rapidly growing research community and widening market for ODRSs, as well as the surging wave of edge intelligence, we find our proposed tutorial a timely opportunity to provide an overview of existing research on this new-generation lightweight recommendation paradigm and an outlook on the future development of ODRSs. This tutorial, or any form of its variation, has not been previously presented in a different venue by any member of the tutorial team. Furthermore, we have conducted a thorough search for relevant ODRSs tutorials with the keyword \textit{recommendation} or \textit{recommender systems} at all the top computer science conferences in the recent five years and only identified one relevant tutorial presented at the IJCAI'20, called Federated Recommender Systems\footnote{https://www.fedai.org/research/conferences/ijcai-2020-tutorial/}. This tutorial introduced the concepts of vertical and horizontal federated RSs, where two case studies on news recommendation and online advertising were presented. However, federated recommendation is only one of the possible technical pathways with on-device training, which in turn is a subset of ODRSs. Our tutorial will not only introduce additional on-device training methods, such as semi-decentralized on-device recommendation and on-device recommender finetuning but will also cover on-device deployment and inference as well as privacy and security mechanisms in ODRSs. To our knowledge, the proposed tutorial will debut the very first comprehensive summary of the fundamentals and recent advances of on-device recommender systems in the research community.

\textbf{Relevance to the Web Conference (WWW).}
Every year, WWW attracts a considerable proportion of high-quality papers and conference attendees working on RSs. The growing significance of RSs is evident, as demonstrated by the dedicated recommendation research track at WWW, as well as the increasing number of research papers and industry participation. For instance, in WWW'23, 19.3\% accepted regular papers (79/409) contained the keyword \textit{recommendation} or \textit{recommender systems}, highlighting the substantial interest in this field. The dense population of experts in relevant areas ensures a vibrant environment for knowledge exchange, constructive discussions, and the exploration of innovative ideas that can shape the future of RSs.

Furthermore, the regular industry sponsors of WWW, including renowned companies like Google, Amazon, Baidu, Ebay, Netflix, and Booking, have dedicated commercial branches focused on recommendation services. Their involvement signifies the practical implications and economic potential of RSs, further solidifying WWW's status as a prominent conference for promoting next-generation ODRSs.

Given the conference's widespread presence of research and industry leaders in the recommendation domain, WWW'24 holds great promise as an ideal platform to disseminate fundamental knowledge, promote recent research outcomes, and foster collaborative efforts to enhance ODRSs. By embracing the wisdom and collective expertise of the conference participants, this tutorial expects to advance the field and address the open challenges associated with on-device recommendation technologies.

\subsection{THE TUTORIAL TEAM}
Prof. Hongzhi Yin and his research group are the pioneers of this emerging research field and have consistently worked on recommender systems for years. Together with co-authors, their work on on-device recommender systems has been published in top-tier venues such as KDD, WWW, SIGIR, AAAI, TKDE, WSDM, TOIS and etc. 

\subsubsection{Brief Bio of Organizers}
\begin{itemize}
    \item \textbf{Prof. Hongzhi Yin} works as an ARC Future Fellow, Full Professor, and Director of the Responsible Big Data Intelligence Lab (RBDI) at The University of Queensland, Australia. He has published 260+ papers with an H-index of 66,  making notable contributions to recommendation systems,  graph learning, decentralized learning, and edge intelligence. His research has won 8 international and national Best Paper Awards, including Best Paper Award (Honorable Mention) at WSDM 2023, Best Paper Award at ICDE 2019, and Best Student Paper Award at DASFAA 2020. He has received the prestigious 2023 AIPS Young Tall Poppy Science Awards, 2022 IEEE Computer Society AI's 10 to Watch, 2021 ARC Future Fellowship, and 2016 ARC DECRA Fellowship. He has been an SPC or area chair for many top conferences, such as WWW, IJCAI, AAAI, KDD, SIGIR, WSDM, ICDE, CIKM, and DASFAA. Prof. Yin has rich lecture experience and taught five relevant courses, such as information retrieval and web search, data mining, social media analytics, and responsible data science. He won the Faculty Teaching and Learning Excellence Award 2022 and the University Teaching and Learning Excellence Award 2022 (finalist). 
In addition, he has delivered 20+ keynotes and tutorials at the top international conferences like DASFAA'23, WWW'22, BESC'22, ADMA'19, WWW'17, and KDD'17.
  \item \textbf{Dr. Tong Chen} is a senior lecturer at The University of Queensland, and an awardee of the 2023 Discovery Early Career Researcher Award from the Australian Research Council (ARC). Dr. Chen's research on lightweight and on-device recommender systems has been published on top-tier international venues such as KDD, SIGIR, WWW, TKDE, WSDM, TNNLS, TOIS, and CIKM. Dr. Chen has ample track records in lecturing, witnessed by his course design and delivery experience in business analytics, teaching experience in social media analytics, as well as invited talks on cutting-edge recommender systems at the DASFAA'23 Tutorial, WWW'22 Tutorial, and ICDM'20 NeuRec Workshop.
    \item \textbf{Mr. Liang Qu} is currently pursuing his Ph.D. under a joint program between The University of Queensland and Southern University of Science and Technology. In 2017, he earned his B.E. in Applied Physics from the South China University of Technology, followed by an M.S. in Computer Science in 2019 from the Harbin Institute of Technology. His research work has been published on top data mining venues such as KDD, SIGIR, WWW, and TOIS. In addition, he has been an PC and/or reviewer for many top venues, such as KDD, WWW, CIKM, and VLDB. His research interest primarily lies in the development of lightweight, privacy-preserving, and trustworthy recommender systems, such as federated recommendation and on-device recommendation.
    \item \textbf{Prof. Bin Cui} is a Cheung Kong Distinguished Professor, Vice Dean of the School of Computer Science at Peking University, and Director of Peking University-Tencent Joint Innovation Laboratory. His research interests include recommendation and search system architectures, query and index techniques, big data management and mining, and distributed machine learning systems. He has served on the Technical Program Committee of various international conferences, including SIGMOD, VLDB, ICDE, WWW, KDD, and as Area Chair of ICDE 2011\&2018, Demo Co-Chair of ICDE 2014, Area Chair of VLDB 2014, PC Co-Chair of APWeb 2015, WAIM 2016 and DASFAA 2020. He serves as Vice Chair of Technical Committee on Database China Computer Federation (CCF) and Trustee Board Member of VLDB Endowment. He is also on the Editorial Board of Distributed and Parallel Databases, Journal of Computer Science and Technology, and SCIENCE CHINA Information Sciences, and was an associate editor of IEEE Transactions on Knowledge and Data Engineering (TKDE) and VLDB Journal. He was awarded Microsoft Young Professorship Award (MSRA 2008), CCF Young Scientist Award (2009), Second Prize of Natural Science Award of MOE China (2014), and appointed as Cheung Kong Distinguished Professor by MOE in 2016.
\end{itemize}

\subsubsection{Relevant Publications by Organizers}
In order to further demonstrate that the presenters are qualified for a high-quality introduction of the ODRSs, below we list the relevant papers on ODRSs published by the presenters.

\begin{itemize}
    \item Deployment and inference for ODRSs \cite{qu2023continuous,xia2022device,chen2021learning,yang2023vqgraph,liang2023learning,qu2023budgeted,xia2023towards,xia2023efficient}
    \item Training for ODRSs \cite{imran2023refrs,long2023decentralized,qu2023semi,long2023model,wang2021fast}
    \item Privacy and security for ODRSs \cite{yuan2023interaction,yuan2023federated,zhang2022pipattack,yuan2023manipulating,zhang2023comprehensive}
\end{itemize}

Overall, this tutorial aims to benefit the participating audience from the following three aspects:
\begin{itemize}
    \item We aim to furnish participants with a comprehensive and current picture of ODRSs, enabling them to grasp the current state-of-the-art technologies and methodologies employed in ODRSs.
    \item We will lay out a systematic categorization of ODRSs for participants, facilitating a structured understanding of the various methods involved. Each category will be explored in detail, discussing the technical aspects that differentiate them. 
    \item We will outline potential future research directions in the ODRS, aiding participants in identifying areas where they can contribute and further the body of knowledge in this field.
\end{itemize}

\section{Style}
This tutorial is delivered as a lecture-style tutorial, which aims to provide a comprehensive introduction to research on ODRSs, from pioneering work to state-of-the-art research, and also discuss future research directions and challenges.

\section{Schedule}\label{sec:content}
The content is planned for 3 hours and consists of five sections. In what follows, we provide an outline of our tutorial.

\noindent\fbox{%
    \parbox{0.47\textwidth}{%
\noindent\textbf{Section 1. Welcome and Introduction (10 mins)}\\
\textbf{Presenter: Prof. Hongzhi Yin} \\
1.1 Overview of Recommender Systems (RSs) \cite{koren2021advances}\\
1.2 On-Device Recommender Systems (ODRSs): Background and Applications \cite{gong2020edgerec,gong2022real,lv2023duet,wang2020next}
    }%
}

\noindent\fbox{%
    \parbox{0.47\textwidth}{%
\noindent\textbf{Section 2. Definition and Taxonomy of ODRSs (20 mins)}\\
\textbf{Presenter: Prof. Hongzhi Yin} \\
2.1 Definition of On-Device Recommendation Tasks\\
2.2 Categorization of Existing ODRSs
    }%
}

\noindent\fbox{%
    \parbox{0.47\textwidth}{%
        \noindent\textbf{Section 3. A Review of ODRSs (110 mins)}\\
        \textbf{Presenters: Dr. Tong Chen and Mr. Liang Qu} \\
3.1 On-Device Deployment and Inference:
\begin{itemize}
\item Binary Code-based Methods \cite{zhang2016discrete,zhang2017discrete}
\item Embedding Sparsification Methods \cite{liu2021learnable,qu2023continuous}
\item Compositional Embedding Methods \cite{wang2020next,shi2020compositional,lian2020lightrec,xia2022device,xia2023efficient}
\item Variable Size Embedding Methods \cite{liu2020automated,chen2021learning,kang2021learning}
\item Sustainable Deployment 
\cite{xia2023towards}
\end{itemize}
3.2 On-Device Training:
\begin{itemize}
	\item Server-coordinated/Federated Learning for On-device Recommendation \cite{muhammad2020fedfast,imran2023refrs,zhang2023lightfr}
	\item Semi-decentralized ODRSs \cite{long2023decentralized,qu2023semi,long2023model}
	\item On-device Recommender Finetuning \cite{wang2021fast,yao2021device,yan2022device}
\end{itemize}
3.3 Privacy and Security:
\begin{itemize}
	\item Privacy Risks and Countermeasures \cite{yuan2023interaction,chai2022efficient,yuan2023federated,zhang2023comprehensive}
	\item Poisoning Attacks and Defense Methods \cite{zhang2022pipattack,wu2022fedattack,yuan2023manipulating}
\end{itemize}
    }%
}

\noindent\fbox{%
    \parbox{0.47\textwidth}{%
        \noindent\textbf{Section 4. Limitations and New Trends (20 mins)}\\
        \textbf{Presenter: Prof. Bin Cui}\\
        4.1 Open Challenges for Existing ODRSs\\
		4.2 Emerging Research Directions
    }%
}

\noindent\fbox{%
    \parbox{0.47\textwidth}{%
        \noindent\textbf{Section 5. Open Discussions (20 mins)}\\
        \textbf{Presenters: Prof. Hongzhi Yin, Dr. Tong Chen, Mr. Liang Qu and Prof. Bin Cui}\\
5.1 Questions and Answers\\
5.2 Reflections, Suggestions, and Link to Our Resources
    }%
}

\section{Audience}
This tutorial targets a diverse audience cohort from both academia and industry, with a background of recommendation or any relevant areas, including but not limited to information retrieval, web mining, and internet-of-things (IoT). 
For prerequisites, basic knowledge of recommender systems is preferred, while the tutorial will also cover all necessary foundations for better audience engagement. After the tutorial, we expect the audience to form an up-to-date picture of different application scenarios of ODRSs, as well as their core technical building blocks. Considering the high accessibility of the conference, we expect around a hundred participants for the tutorial.

\section{TUTORIAL MATERIALS}
Upon acceptance of the tutorial, the slides and video recordings will be made available to all attendees on our tutorial website two weeks before the scheduled conference date.

\section{VIDEO TEASER}
The video teaser is available at \url{https://bit.ly/odrs}.

\section*{Acknowledgment}
This work is supported by the Australian Research Council under the streams of Future Fellowship (No. FT210100624), Discovery Project (No. DP190101985, DP240101108, DP240101814), and Discovery Early Career Researcher Award (No. DE230101033).


\end{document}